\documentclass[%
 reprint,
 amsmath,amssymb,
 aps,
]{revtex4-1}

\usepackage{graphicx}
\usepackage{graphics}
\usepackage{dcolumn}
\usepackage{bm}
\usepackage{float}
\usepackage{caption}
\usepackage{subfigure}
\usepackage{amsmath}
\usepackage{mathrsfs}
\usepackage{amssymb}
\usepackage{chemarrow}
\usepackage{hyperref}

\begin{document}

\preprint{}

\title{Large Deviations for Chiral Transition through Path Integral}

\author{Li Wan}
\email{lwan@wzu.edu.cn}
\author{Yunmi Huang}
\affiliation{Department of Physics, Wenzhou University, Wenzhou 325035, P. R. China}
\author{Xiaoshan Zhu}
\affiliation{Department of Electrical and Biomedical Engineering, University of Nevada, Reno, USA}
\date{\today}

\begin{abstract}
An noise-induced mechanism has been revealed by some authors recently for the homochirality in a chiral system. Motivated by such stochastic process, we study the noise-induced transition in the system. The chiral transition, say the transition between chiral states, is impossible in the deterministic view if the homochirality has been converged, but is a rare event when noise is involved. We study the rare event by using the large deviation theory (LDT) and figure out the LDT rate functional for the transition probability through the Doi-Peliti second quantization path integral method. In order to check the correctness of our work, we have derived the Langevin equations through the path integral method by using the Hubbard-Stratonovich transformation, and recover the equations reported by other authors through a different method.
\begin{description}
\item[PACS numbers]
05.10.Gg, 05.40.-a, 82.20.-w
\item[Keywords]
homochirality,  path integral, large deviation theory, noise, chiral transition
\end{description}
\end{abstract}

\maketitle

\section{Introduction}
Homochirality is one of the universal features for chiral systems. It has been observed that in biological systems every naturally occurring amino acid is left-handed and every sugar is right-handed~\cite{blackmond, Gleiser}. Such symmetry breaking is also well known in physics and chemical reactions~\cite{Soai,Saito,Kipping,Kipping1,Kondepudi,Viedma,Noorduin,Viedma1,Tsogoeva}. The homochirality is referred to as such symmetry breaking, meaning that a chiral system converges to one of its chiral states in the time evolving. The process for the homochirality is dependent on initial conditions, such as the initial majority and functions of catalysis~\cite{Saito}. The mechanism for the homochirality has not been fully understood yet. Despite of the fact of the unknown mechanism, it has been well accepted that the chiral state for the homochirality is a fixed state after the convergence of the chiral system, which is known as the deterministic view. In such deterministic view, the chiral transition in the system, say the transition between chiral states of the system, is impossible. Recently, a stochastic model has been proposed, showing that noise can induce the homochirality for the chiral system~\cite{Jafarpour}. Here, the noise is originated from the density fluctuation of the chiral substances, and is influenced by the size of the system. With the noise involved in the system, the chiral transition becomes possible and is a rare event for the system. In this paper, we use the large deviation theory (LDT) to study the rare event and figure out the LDF rate functional for the transition probability~\cite{Touchette}.\\

To understand the homochirality, Frank proposed the first model in which an achiral molecule $A$ turns into a chiral product $D$ (or $L$) if $A$ comes into contact with $D$(or $L$)~\cite{Frank}. $D$ and $L$ are enantiomers of a chiral molecule. The reaction process can be represented by $A+D\rightarrow D$~ and $A+L\rightarrow L$. This process is understood to be a linear autocatalytic process or the first order process. In the model, Frank also suggested a process of mutual antagonism in which the two opposite enantiomers of $D$ and $L$ annihilate when they come into contact. That is $D+L\rightarrow 0$. By solving the deterministic equations established from the Frank's model, two steady states can be obtained. With an initial chiral bias, the homocharility then is achieved.\\

The Frank's model has been developed with many other mechanisms included within the deterministic frames~\cite{Uwaha,Uwaha1,Saito1,Saito2,McBride}. In the developed models, the spontaneous process and the nonlinear autocatalytic process have been fully studied. The spontaneous process is $A\rightarrow D$ and $A \rightarrow L$, while the nonlinear autocatalytic process is $A+2D\rightarrow 3D$ and $A+2L\rightarrow 3L$. The nonlinear autocatalytic process is also called the second order process. To get the homochirality, one should amplify the enantiomeric excess (EE) as the first step and then drive the system to reach the homochirality as the second step~\cite{Saito}. The nonlinear process has been found to play an important role in the amplification of EE. For the second step, the recycling process is essential, which is $D\rightarrow A$ and $L\rightarrow A$. The mechanism for the recycling process is still not clear, which can not be simply understood as the reverse reaction of the spontaneous process. In the deterministic models, the chiral state for the homochirality is a fixed state, and it is no way for the system to transit between the chiral states after the convergence of the chiral state.\\

Besides the deterministic models, stochastic models provide a different understanding for the homochirality~\cite{Lente,Saito3}. The stochastic view is originated from the fluctuation of the final EE value in each experimental trial~\cite{Singleton, Soai1,Gridnev,Kawasaki}. Recently, an noise-induced mechanism for the homochirality has been proposed~\cite{Jafarpour}. The noise mentioned in the model is originated from the density fluctuation of the chiral substances, and is influenced by the size of the system. In the stochastic model, the spontaneous process and the linear autocatalytic process combining the recycling process are used to establish stochastic differential equations, through which the homochiraltiy can be obtained. According to the model, the nonlinear autocatalytic process is unnecessary for the achievement of the homochirality, which is opposite to the deterministic model. The homochirality and the racemic state can be manipulated by adjusting the size of the system in the model. With the noise present in the system, it is natural for us to think that the chiral transition in the system becomes possible and is a rare event for the system. To identify the transition probability, it is necessary for us to get the LDT rate functional for the rare event, which is the main goal in this work.\\

\section{transition probability in a path integral form}
We study the following reaction processes~\cite{Jafarpour},
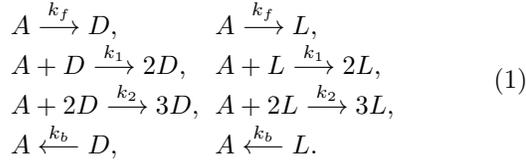
\begin{align}
\label{reaction}
\begin{array}{ll}
A \stackrel{k_f}{\longrightarrow} D, & A \stackrel{k_f}{\longrightarrow} L, \\
A+D \stackrel{k_1}{\longrightarrow} 2D, & A+L \stackrel{k_1}{\longrightarrow} 2L, \\
A+2D \stackrel{k_2}{\longrightarrow} 3D, & A+2L \stackrel{k_2}{\longrightarrow} 3L, \\
A \stackrel{k_b}{\longleftarrow} D, & A \stackrel{k_b}{\longleftarrow} L.
\end{array}
\end{align}
The first line of eq.(\ref{reaction}) is for the spontaneous process. The second and the third lines are for the linear process and the nonlinear process respectively. The last line is for the recycling process. It has been revealed that the nonlinear process is not essential for the homochirality in the stochastic view. But we include the nonlinear process in this study for generality, considering that the nonlinear process play an important role in the deterministic view. $k_f$, $k_1$, $k_2$ and $k_b$ are the reaction rates for their corresponding processes. The arrows in eq.(\ref{reaction}) mean the reaction directions for the processes. The probability for the system having the particle number $a$ for the substance $A$, number $d$ for the substance $D$ and number $l$ for the substance $L$ is denoted by $P(a,d,l)$. The state of the system with the particle numbers $a$, $d$ and $l$ is represented by a vector $|a,d,l>$. The reaction radius of one particle is $r_0$ and the probability for one particle meeting another particle in the reaction radius is $\frac{4\pi r_0^3}{3\Omega}$ with the $\Omega$ the total volume of the system. For notational simplicity, the meeting probability is written as $\frac{1}{\Omega}$. For simplicity, we assume the reaction processes of the substances in the system are not limited by their diffusion processes. The distribution of the reaction substances is uniform in the volume and is independent on position.  So, the equation for the $P(a,d,l)$ is
\begin{widetext}
\begin{align}
\label{probabilityequ}
\frac{\partial P(a,d,l)}{\partial t}&=P(a+1,d-1,l)[\frac{k_1}{\Omega}(a+1)(d-1)+\frac{k_2}{\Omega^2}(a+1)(d-1)(d-2)+k_f(a+1)]\nonumber \\
&+P(a+1,d,l-1)[\frac{k_1}{\Omega}(a+1)(l-1)+\frac{k_2}{\Omega^2}(a+1)(l-1)(l-2)+k_f(a+1)]\nonumber \\
&+k_b(d+1)P(a-1,d+1,l)+k_b(l+1)P(a-1,d,l+1)\nonumber \\
&-P(a,d,l)[\frac{k_1}{\Omega}a(d+l)+\frac{k_2}{\Omega^2}a(d^2+l^2-d-l)+2k_f a+k_b(d+l)].
\end{align}
\end{widetext} 
In the following, we use the Doi-Peliti second quantization path integral method to find the transition probability~\cite{Doi,Peliti}. The wave function of the system then can be defined as
\begin{align}
\label{wavefunc}
\Phi(t)=\sum \limits_{a,d,l=0}^{\infty}P(a,d,l)|a,d,l>.
\end{align}
Now we introduce a creation operator $\hat{a}^+$ and an annihilation operator $\hat{a}$ for the substance $A$ ($\hat{d}^+$ and $\hat{d}$~for~ $D$~, $\hat{l}^+$ and $\hat{l} $~for $L$). The operators have following definitions
\begin{align}
&\hat{a}|n+1>=(n+1)|n>,~~\hat{a}^+|n>=|n+1>,\nonumber \\
&[\hat{a},\hat{a}^+]=1,~~ [\hat{a},\hat{a}]=[\hat{a}^+,\hat{a}^+]=0,~~\hat{a}^+\hat{a}|n>=n|n>.
\end{align} 
By replacing the $a$ in the above equations with $d$ or $l$ , we get similar definitions for the operators $\hat{d}$, $\hat{d}^+$, $\hat{l}$ and $\hat{l}^+$. Thus, by using the operators, eq.(\ref{probabilityequ}) can be represented by
\begin{align}
\frac{\partial \Phi(t)}{\partial t}=-\hat{H} \Phi(t),
\end{align}
with $\hat{H}$ the Hamiltonian reading
\begin{align}
\label{Hamil}
\hat{H}=-\frac{k_1}{\Omega}A-\frac{k_2}{\Omega^2}B-k_fC-k_bD.
\end{align}
Here, we have
\begin{align}
\label{Hamil1}
& A=\hat{a} \hat{d}^{+}\hat{d}^+\hat{d}+\hat{a} \hat{l}^{+}\hat{l}^+\hat{l}-\hat{a}^+\hat{a}(\hat{d}^+\hat{d}+\hat{l}^+\hat{l}),\nonumber \\
& B=\hat{a} (\hat{d}^{+})^3(\hat{d})^2+\hat{a} (\hat{l}^{+})^3(\hat{l})^2-\hat{a}^+\hat{a}(\hat{d}^+\hat{d}^+\hat{d}\hat{d}+\hat{l}^+\hat{l}^+\hat{l}\hat{l}),\nonumber \\
& C=\hat{a} \hat{d}^{+}+\hat{a} \hat{l}^{+}-2\hat{a}^+\hat{a},\nonumber \\
& D=\hat{a}^+ \hat{d}+\hat{a}^+ \hat{l}-\hat{d}^+\hat{d}-\hat{l}^+\hat{l}.
\end{align}
Derivations for the above equations can be found in the supplementary material of this paper. 

\subsection{vector space}
We have defined the vector $|m_a, m_d, m_l>$ for the state of the system with the particle numbers $m_a$, $m_d$ and $m_l$ of the substances $A$, $D$ and $L$ respectively. And then we define a vector for a coherent state by
\begin{align}
|Z(z_a, z_d, z_l)>=e^{z_a\hat{a}^+}e^{z_d\hat{d}^+}e^{z_l\hat{l}^+}|0,0,0>.
\end{align}
Especially, we simplify the notation of vector $|Z(1, 1, 1)>$ by
\begin{align}
|\mathcal{P}>=|Z(1, 1, 1)>.
\end{align}
The coherent states have following relations
\begin{align}
&\hat{a}|Z(z_a, z_d, z_l)>=z_a|Z(z_a, z_d, z_l)> \nonumber \\
&\hat{d}|Z(z_a, z_d, z_l)>=z_d|Z(z_a, z_d, z_l)> \nonumber \\
&\hat{l}|Z(z_a, z_d, z_l)>=z_l|Z(z_a, z_d, z_l)>.
\end{align}
The inner products of the vectors are
\begin{align}
\label{innerprod}
&<m_a, m_d,m_l|n_a,n_d,n_l>=m_a!m_d!m_l!\delta_{m_a, n_a}\delta_{m_d, n_d}\delta_{m_l, n_l},\nonumber \\
&<m_a, m_d,m_l|\mathcal{P}>=1,\nonumber \\
&<m_a,m_d,m_l|Z(z_a, z_d, z_l)>=(z_a)^{m_a}(z_d)^{m_d}(z_l)^{m_l}.
\end{align}
The statistical average of any operator $\hat{O}$ can be obtained from the second line of eq.(\ref{innerprod}) as
\begin{align}
<O>=<\mathcal{P}|\hat{O}|\Phi>.
\end{align}
If $z_a$,$z_d$ and $z_l$ have following forms
\begin{align}
\label{zadl}
z_a=\mu_ae^{-i\theta_a},~z_d=\mu_de^{-i\theta_d},~z_l=\mu_le^{-i\theta_l},
\end{align}
we can define a reduced vector for the coherent state
\begin{align}
|\tilde{Z}(\theta_a,\theta_d,\theta_l)>=|Z(e^{-i\theta_a},e^{-i\theta_d},e^{-i\theta_l})>.
\end{align}

\subsection{path integral form}
The unity in constructing the path-integral expression is decomposed with the coherent stats by
\begin{align}
\label{unity}
\mathbf{1}=\mathcal{I}_a\mathcal{I}_d\mathcal{I}_le^{-(\mu_a+\mu_d+\mu_l)}|Z(z_a,z_d,z_l)><\tilde{Z}(\theta_a,\theta_d,\theta_l)|
\end{align}
with 
\begin{align}
\label{intsymb}
\mathcal{I}_a=\int_0^{\infty}d\mu_a\int_{-\pi}^{\pi}\frac{d\theta_a}{2\pi}
\end{align}
to simplify the notation of integral~\cite{Itakura}. Here, $z_a$,~$z_d$ ~and~ $z_l$ have the forms shown in eq.(\ref{zadl}). At the initial time  $t=0$ we denote the particle numbers by $m_a^i$, ~$m_d^i$ ~ and $m_l^i$ for the substances $A$, $D$ and $L$ respectively. Here, the superscript $i$ means the $''initial''$. Similarly, the particle numbers at the final time $t=T$ are denoted by $m_a^f$, ~$m_d^f$ ~ and $m_l^f$ correspondingly with the superscript $f$ meaning $''final''$. Then the vector for the state at the initial time is $|m_a^i,m_d^i,m_l^i>$, and the vector for the final state is $|m_a^f,m_d^f,m_l^f>$. The transition probability from the initial state to the final state is
\begin{align}
&P(m_a^f,m_d^f,m_l^f;T|m_a^i,m_d^i,m_l^i;0)\nonumber \\
=&\frac{1}{m_a^f!m_d^f!m_l^f!}<m_a^f,m_d^f,m_l^f|e^{-\hat{H}T}|m_a^i,m_d^i,m_l^i>.
\end{align}
Now we insert the unity eq.(\ref{unity}) between time slices along the transition path, and the transition probability is expressed in the path-integral form as
\begin{widetext}
\begin{align}
\label{PI}
P(m_a^f,m_d^f,m_l^f;T|m_a^i,m_d^i,m_l^i;0)=\lim \limits_{\tau\rightarrow 0}(\prod \limits_{t=\tau}^{T-\tau}\mathcal{I}_a^t\mathcal{I}_d^t\mathcal{I}_l^t)\frac{1}{m_a^f!m_d^f!m_l^f!}(m_a^f)^{m_a^f}(m_d^f)^{m_d^f}(m_l^f)^{m_l^f}e^{-(m_a^f+m_d^f+m_l^f)} \times e^{-\mathcal{S}},
\end{align}
with the action
\begin{align}
\label{S}
\mathcal{S}=\sum \limits_{t=\tau}^T[H(\tilde{Z}_t^*, Z_{t-\tau})\tau+i\theta_a^t(\mu_a^t-\mu_a^{t-\tau})+i\theta_d^t(\mu_d^t-\mu_d^{t-\tau})+i\theta_l^t(\mu_l^t-\mu_l^{t-\tau})].
\end{align}
\end{widetext}
Note that in eq.(\ref{PI}) the factor of $(m_a^f)^{m_a^f}$ is $m_a^f$ in the $(\cdots)$ raised to the power of the superscript $m_a^f$. We need to specify that in order to get the terms $H(\tilde{Z}_t^*, Z_{t-\tau})$ in the $\mathcal{S}$, the operators in the Hamiltonian $\hat{H}$ should have the order of eq.(\ref{Hamil1}) with the creation operators placed at the left side of the annihilation operators for each substance, say $\hat{a}^+\hat{a}$, $\hat{d}^+\hat{d}$ and $\hat{l}^+\hat{l}$ in order. $\tilde{Z}_t^*$ means that $\hat{a}^+$, $\hat{d}^+$ and $\hat{l}^+$ in the Hamiltonian $\hat{H}$ are replaced by $e^{i \theta_a^t}$, $e^{i \theta_d^t}$ and $e^{i \theta_l^t}$ at the time instant $t$. $Z_{t-\tau}$ means that $\hat{a}$, $\hat{d}$ and $\hat{l}$ in the  Hamiltonian $\hat{H}$ are replaced by $\mu_a^{t-\tau}e^{-i \theta_a^{t-\tau}}$, $\mu_d^{t-\tau}e^{-i \theta_d^{t-\tau}}$ and $\mu_l^{t-\tau}e^{-i \theta_l^{t-\tau}}$ at the time instant $t-\tau$. In this way, we get the  $H(\tilde{Z}_t^*, Z_{t-\tau})$.\\

By integrating the functions at the initial time and the final time, we obtain delta functions showing the following constraint conditions 
\begin{align}
\label{restrictcond}
&\mu_a^i=m_a^i,\mu_d^i=m_d^i,\mu_l^i=m_l^i,\nonumber \\
&\mu_a^T=m_a^f,\mu_d^T=m_d^f,\mu_l^T=m_l^f,
\end{align}
for eq.(\ref{PI}). The above eq.(\ref{restrictcond}) shows that the variables $\mu_a$, $\mu_d$ and $\mu_l$ actually have the physical meanings of the particle numbers. Details for the derivation can be found in the supplementary material.\\

Now we introduce the density of the substances in the volume $\Omega$ of the system by
\begin{align}
\label{varb1}
\eta_t=\frac{\mu_a^t}{\Omega},~~\kappa_t=\frac{\mu_d^t}{\Omega},~~\zeta_t=\frac{\mu_l^t}{\Omega}.
\end{align} 
To keep the Jacobian determinant as the unit for the integral $\mathcal{I}^t$ in eq.(\ref{intsymb}), we make variable transformations by
\begin{align}
\label{varb2}
\alpha_t=\theta_a^t\Omega,~~\beta_t=\theta_d^t\Omega,~~\gamma_t=\theta_l^t\Omega.
\end{align} 
Thus, the integral $\mathcal{I}^t$ reads
\begin{align}
&\mathcal{I}_a^t=\int_0^{\infty} d \eta_t\int_{-\infty}^{\infty}\frac{d \alpha_t}{2\pi},\nonumber \\
&\mathcal{I}_d^t=\int_0^{\infty} d \kappa_t\int_{-\infty}^{\infty}\frac{d \beta_t}{2\pi},\nonumber \\
&\mathcal{I}_l^t=\int_0^{\infty} d \zeta_t\int_{-\infty}^{\infty}\frac{d \gamma_t}{2\pi},
\end{align}
under the consideration that the volume $\Omega$ is large enough.

\section{Large Deviations}
In eq.(\ref{PI}), we approximate the factor
\begin{align}
\frac{1}{m_a^f!m_d^f!m_l^f!}(m_a^f)^{m_a^f}(m_d^f)^{m_d^f}(m_l^f)^{m_l^f}e^{-(m_a^f+m_d^f+m_l^f)}\approxeq 1
\end{align}
for $m_a^f \gg 1$, $m_d^f \gg 1$ and $m_l^f \gg 1$. Based on eq.(\ref{varb1}) and eq.(\ref{varb2}), the obtained $H(\tilde{Z}_t^*,Z_{t-\tau})$ reads
\begin{align}
\label{Hamil2}
&H(\tilde{Z}_t^*,Z_{t-\tau})=\frac{\lambda_1}{2\Omega}(\alpha_{t-\tau}-\beta_t)^2+\frac{\lambda_2}{2\Omega}(\alpha_{t-\tau}-\gamma_t)^2\nonumber \\
&+i\lambda_3(\alpha_{t-\tau}-\beta_t)+i\lambda_4(\alpha_{t-\tau}-\gamma_t),
\end{align}
with
\begin{align}
&\lambda_1=k_1\eta_{t-\tau}\kappa_{t-\tau}+k_2\eta_{t-\tau}\kappa_{t-\tau}^2+k_f\eta_{t-\tau}+k_b\kappa_{t-\tau},\nonumber \\
&\lambda_2=k_1\eta_{t-\tau}\zeta_{t-\tau}+k_2\eta_{t-\tau}\zeta_{t-\tau}^2+k_f\eta_{t-\tau}+k_b\zeta_{t-\tau}, \nonumber \\
& \lambda_3=k_1\eta_{t-\tau}\kappa_{t-\tau}+k_2\eta_{t-\tau}\kappa_{t-\tau}^2+k_f\eta_{t-\tau}-k_b\kappa_{t-\tau},\nonumber \\
& \lambda_4=k_1\eta_{t-\tau}\zeta_{t-\tau}+k_2\eta_{t-\tau}\zeta_{t-\tau}^2+k_f\eta_{t-\tau}-k_b\zeta_{t-\tau}.
\end{align}
So the action $\mathcal{S}$ takes the form of 
\begin{align}
\label{S1}
&\mathcal{S}=\tau\sum \limits_{t=\tau}^T[\frac{\lambda_1}{2\Omega}(\alpha_{t}-\beta_t)^2+\frac{\lambda_2}{2\Omega}(\alpha_{t}-\gamma_t)^2\nonumber \\
&+i\lambda_3(\alpha_{t}-\beta_t)+i\lambda_4(\alpha_{t}-\gamma_t)\nonumber \\
&+i\lambda_5\alpha_t+i\lambda_6\beta_t+i\lambda_7\gamma_t]
\end{align}
with
\begin{align}
\lambda_5=\frac{\eta_t-\eta_{t-\tau}}{\tau}, ~~\lambda_6=\frac{\kappa_t-\kappa_{t-\tau}}{\tau}, ~~\lambda_7=\frac{\zeta_t-\zeta_{t-\tau}}{\tau}.
\end{align}
$\alpha_{t-\tau}$, $\beta_{t-\tau}$ and $\gamma_{t-\tau}$ in eq.(\ref{Hamil2}) have been replaced by $\alpha_{t}$, $\beta_{t}$ and $\gamma_{t}$ respectively in eq.(\ref{S1}) with the consideration that $\tau$ is infinitesimal.\\

Completing the squares of $\alpha_t$ and $\beta_t$ in the action $\mathcal{S}$ and then integrating out $\alpha_t$ and $\beta_t$ through $\int_{-\infty}^{\infty}\frac{d \alpha_t}{2\pi}\int_{-\infty}^{\infty}\frac{d \beta_t}{2\pi}$ in  $\mathcal{I}_a^t\mathcal{I}_d^t$, we get remained terms of $\mathcal{S}$ as
\begin{align}
\label{Sremained}
&\mathcal{S}_{rem}=\tau \sum_{t=\tau}^T [i\gamma_t[\lambda_5+\lambda_6+\lambda_7]\nonumber \\
&+\Omega[\frac{\lambda_1(\lambda_7-\lambda_4)^2+\lambda_2(\lambda_6-\lambda_3)^2}{2\lambda_1\lambda_2}]].
\end{align}
Further, we integrate out $\gamma_t$ in eq.(\ref{Sremained}) through $\int_{-\infty}^{\infty}\frac{d \gamma_t}{2\pi}$ in the $\mathcal{I}_l^t$, we get 
\begin{align}
\label{conservation}
\lambda_5+\lambda_6+\lambda_7=0,
\end{align}
meaning the conservation law of particle numbers. The transition probability eq.(\ref{PI}) then is reduced to be
\begin{widetext}
\begin{align}
\label{reducedPI}
P(m_a^f,m_d^f,m_l^f;T|m_a^i,m_d^i,m_l^i;0)\simeq \lim \limits_{\tau\rightarrow 0}(\prod \limits_{t=\tau}^{T-\tau}\int_0^{\infty} d\eta_t\int_0^{\infty} d\kappa_t\int_0^{\infty} d\zeta_t) \times e^{-\tau \sum_{t=\tau}^T \Omega[\frac{\lambda_1(\lambda_7-\lambda_4)^2+\lambda_2(\lambda_6-\lambda_3)^2}{2\lambda_1\lambda_2}]}.
\end{align}
\end{widetext}

We choose a small bundle of paths $[\eta,\kappa,\zeta]$ for the transition, and assume the probability density for the paths in the transition has a large deviation property by
\begin{align}
P([\eta,\kappa,\zeta])\simeq e^{-\Omega \mathcal{F}[\eta,\kappa,\zeta]}
\end{align}
with $\mathcal{F}[\eta,\kappa,\zeta]$ the LDT rate functional~\cite{Touchette}. Based on the transition probability eq.(\ref{reducedPI}), we arrive at the LDT rate functional
\begin{align}
\label{LDF1}
\mathcal{F}[\eta,\kappa,\zeta]=\tau \sum_{t=\tau}^T [\frac{\lambda_1(\lambda_7-\lambda_4)^2+\lambda_2(\lambda_6-\lambda_3)^2}{2\lambda_1\lambda_2}]
\end{align}
expressed in the time-slicing form. The rate functional can be expressed in the time-continuous form by
\begin{align}
\label{LDF2}
\mathcal{F}[\eta,\kappa,\zeta]=\int _0^T dt [\frac{\lambda_1(\lambda_7-\lambda_4)^2+\lambda_2(\lambda_6-\lambda_3)^2}{2\lambda_1\lambda_2}].
\end{align}
Now $\lambda_{i(i=1,2,3,4,5,6,7)}$ in eq.(\ref{LDF2}) are functions of time $t$ continuously, instead of the quantities in eq.(\ref{LDF1}) discretely. For any physical quantity $O([\eta, \kappa, \zeta])$, the statistic average of the quantity over all possible paths in the transition can be obtained by
\begin{align}
<O(\eta, \kappa, \zeta)>\simeq & \lim \limits_{\tau\rightarrow 0}(\prod \limits_{t=\tau}^{T-\tau}\int_0^{\infty} d\eta_t\int_0^{\infty} d\kappa_t\int_0^{\infty} d\zeta_t)\nonumber \\
&\times e^{-\Omega \mathcal{F}[\eta,\kappa,\zeta]}\times O([\eta, \kappa, \zeta])
\end{align}
under the condition of eq.(\ref{conservation}). Eq.(\ref{LDF1}) and eq.(\ref{LDF2}) are the main result in this work.

\section{Langevin equation}
The conservation law of eq.(\ref{conservation}) obtained from eq.(\ref{PI}) proves the correctness of our derivation. In this section, we will provide one more proof for the correctness. Basically, we will derive Langevin equations from eq.(\ref{PI}) by using the Hubbard-Stratonovich transformation (HST) and compare the equations to the result reported by other authors through Fokker-Planck equation. We will find that the equations in our work can recover the reported result. \\

The component of $\mathcal{S}$  at a time instant $t$ in eq.(\ref{S1}) is denoted by $\mathcal{S}_t$ for notational simplicity. The $\mathcal{S}_t$ is quadratic, and can be written in a matrix form, reading
\begin{align}
-\mathcal{S}_t&=[i\alpha_t ~~i\beta_t~~i\gamma_t]\left[
\begin{array}{ccc}
\frac{\tau(\lambda_1+\lambda_2)}{2\Omega} &-\frac{\tau \lambda_1}{2\Omega}
&-\frac{\tau \lambda_2}{2\Omega}  \\[2mm]
-\frac{\tau \lambda_1}{2\Omega} &\frac{\tau \lambda_1}{2\Omega} &0  \\[2mm]
-\frac{\tau \lambda_2}{2\Omega} &0 &\frac{\tau \lambda_2}{2\Omega} 
\end{array}
\right]
\left[
\begin{array}{c}
i\alpha_t  \\
i\beta_t  \\
i\gamma_t   
\end{array}
\right]\nonumber \\
&=\frac{1}{2}s\cdot V \cdot s^{tans},
\end{align}
with 
\begin{align}
V=\left[
\begin{array}{ccc}
\frac{\tau(\lambda_1+\lambda_2)}{\Omega} &-\frac{\tau \lambda_1}{\Omega}
&-\frac{\tau \lambda_2}{\Omega}  \\[2mm]
-\frac{\tau \lambda_1}{\Omega} &\frac{\tau \lambda_1}{\Omega} &0  \\[2mm]
-\frac{\tau \lambda_2}{\Omega} &0 &\frac{\tau \lambda_2}{\Omega} 
\end{array}
\right]
\end{align}
and
\begin{align}
s=[i\alpha_t ~~i\beta_t~~i\gamma_t].
\end{align}
Here, the superscript $trans$ means the transpose operator for the vector $s$. We introduce auxiliary variables $\varpi_{i=\alpha, \beta, \gamma}$ and then apply the HST
\begin{align}
\int \prod _i d \varpi_i e^{-\frac{1}{2}\sum _{ij} \varpi_i (V^{-1}_{ij})\varpi_j+\sum_i \varpi_i s_i}=cont \times e^{\frac{1}{2}s\cdot V \cdot s^{trans}}
\end{align}
to make the transformation. Here, $\varpi$ is the noise for the stochastic process with $<\varpi_i>=0$ and $<\varpi_i(t)\varpi_j(t')>=V_{ij}\delta(t-t')$. Thus, the $\mathcal{S}$ in eq.(\ref{PI}) is expressed in the terms of $\varpi$ by using the HST. Then the integration of $e^{\sum_i \varpi_i s_i}$ respecting to $\alpha_t$, $\beta_t$ and $\gamma_t$ in eq.(\ref{PI}) yields a product of $\delta$ functions. The $\delta$ functions actually show the Langevin equations including the noise terms of $\varpi$. Before we give the Langevin equations, we go further by introducing variables $\iota$
\begin{align}
\frac{1}{\sqrt{\tau}}\cdot
\left[
\begin{array}{c}
\varpi_{\alpha}  \\
\varpi_{\beta}  \\
\varpi_{\gamma}   
\end{array}
\right]=Q\cdot
\left[
\begin{array}{c}
\iota_{\alpha}  \\
\iota_{\beta}  \\
\iota_{\gamma}   
\end{array}
\right]
\end{align}
to transform the noise terms. Here, $Q$ is from the Cholesky decomposition by $V=QQ^T$. Finally, we arrive at the Langevin equations
\begin{widetext}
\begin{align}
\label{Langevinequ}
\dot{\eta}=&-k_1\eta(\kappa+\zeta)-k_2\eta(\kappa^2+\zeta^2)-2k_f\eta+k_b(\kappa+\zeta)+\sqrt{\frac{(\lambda_1+\lambda_2)}{\Omega}}\cdot \iota_{\alpha}\nonumber \\
\dot{\kappa}=& k_1\eta\kappa+k_2\eta\kappa^2+k_f\eta-k_b\kappa-\sqrt{\frac{\lambda_1^2}{\Omega(\lambda_1+\lambda_2)}}\cdot \iota_{\alpha}+\sqrt{\frac{ \lambda_1\lambda_2}{\Omega(\lambda_1+\lambda_2)}}\cdot \iota_{\beta} \nonumber \\
\dot{\zeta}=& k_1\eta\zeta+k_2\eta\zeta^2+k_f\eta-k_b\zeta-\sqrt{\frac{\lambda_2^2}{\Omega(\lambda_1+\lambda_2)}}\cdot\iota_{\alpha}-\sqrt{\frac{\lambda_1\lambda_2}{\Omega(\lambda_1+\lambda_2)}}\cdot\iota_{\beta}.
\end{align}
\end{widetext}
The three equations in eq.(\ref{Langevinequ}) actually satisfy $\dot{\eta}+\dot{\kappa}+\dot{\zeta}=0$, showing the conservation law of the particle numbers. The $\iota$ now can be interpreted as Gaussian noise with  
\begin{align}
<\iota_i>=0;~~<\iota_i(t)\iota_j(t')>=\delta_{i,j}\delta(t-t').
\end{align}
If we set $k_2=0$ to drop off the nonlinear process and apply the rule for summing Gaussian variables (i.e. $a \iota_i+b\iota_j=\sqrt{a^2+b^2}\iota$ where $a$ and $b$ are generic functions), we recover eq.(S10) in the supplementary material of Ref.(\cite{Jafarpour}) from eq.(\ref{Langevinequ}), proving the correctness of our work. Details can be found in the supplementary material of this work.\\

The noise considered in the model for the homochirality is from the density fluctuation of the substances, which is influenced by the volume of the system. If the volume of the system approaches infinity, the density fluctuation goes to zero and the noise terms disappear. The Langevin equations then are reduced to deterministic equations.

\section{conclusion}
In this work, we study a stochastic model for homochirality. In the model, the spontaneous process, the linear process, the nonlinear process and the recycling process are considered. With the noise introduced in the system, the transition between the chiral states becomes possible. The probability for the transition is evaluated by the LDT rate functional. We have derived the rate functional through Doi-Peliti second quantization path integral in this paper. In the derivation, the conservation law of the particle numbers is obtained, showing the correctness of our derivation. By using the Hubbard-Stratonovich transformation, we get the Langevin equations for the stochastic process, and recover the equations reported by other authors through a different method, showing that our result is reliable. The LDT rate functional obtained in this paper can be applied for further study of the homochiraltiy, such as non-equilibrium properties, rare events for chiral transition. \\


\end{document}